\documentclass[twocolumn,showpacs]{revtex4}

\usepackage{dcolumn}
\usepackage{latexsym}
\usepackage{amssymb}
\usepackage{graphicx}
\usepackage{bm}
\newcommand{\DE}{{\Delta E}}
\newcommand{\Eb}{\mbox{\boldmath$\mathcal{E}$}}
\newcommand{\epsb}{\mbox{\boldmath$\epsilon$}}
\begin{document}

\title{Microwave traps for cold polar molecules}
\author{D. DeMille, D.R. Glenn, and J. Petricka}
\affiliation{Physics Department, Yale University, P.O. Box 208120,
New Haven, CT 06520 USA}

\date{\today}

\begin{abstract}
We discuss the possibility of trapping polar molecules in the
standing-wave electromagnetic field of a microwave resonant
cavity.  Such a trap has several novel features that make it very
attractive for the development of ultracold molecule sources.
Using commonly available technologies, microwave traps can be
built with large depth (up to several Kelvin) and acceptance
volume (up to several cm$^3$), suitable for efficient loading with
currently available sources of cold polar molecules.  Unlike most
previous traps for molecules, this technology can be used to
confine the strong-field seeking absolute ground state of the
molecule, in a free-space maximum of the microwave electric field.
Such ground state molecules should be immune to inelastic
collisional losses.  We calculate elastic collision cross-sections
for the trapped molecules, due to the electrical polarization of
the molecules at the trap center, and find that they are
extraordinarily large.  Thus, molecules in a microwave trap should
be very amenable to sympathetic and/or evaporative cooling. The
combination of these properties seems to open a clear path to
producing large samples of polar molecules at temperatures much
lower than has been possible previously.
\end{abstract}

\maketitle

%
%
%
%
\section{Introduction}
\label{sec1} A growing interest has developed in extending the
achievements of atomic cooling and trapping to molecules
\cite{coldmolreviews}, particularly \textit{polar} molecules.  The
extremely large electric polarizability of such species enables
access to strikingly new regimes and phenomena.  For example,
polar molecules subjected to strong electric fields have large
laboratory-frame electric dipole moments, and thus can interact
via the very strong, long-range dipole-dipole interaction.  It has
been argued that this can allow trapped polar molecules to be used
as the qubits of a scalable quantum computer \cite{DeMilleQC}.  In
addition, these interactions could make new types of
highly-correlated quantum many-body states accessible, such as
BCS-like superfluids \cite{ShlyapnikovBCS}, supersolid and
checkerboard states, two-dimensional Bose metals
\cite{GoralpolarBEC}, or effectively limited-dimensional gases
\cite{manybodyrefs}.   Other applications include the possibility
to study ultracold chemical reactions
\cite{Dalgarnocoldchemistry}, which might be controlled using
electric fields \cite{Bohnfieldlinked}. Finally, the increased
spectroscopic precision associated with long observation times,
combined with the enhanced sensitivity to certain types of
perturbations, could allow the sensitivity of tests of fundamental
symmetries to be increased to unprecedented levels
\cite{MishaKreview}.

At present, there remains no demonstrated technique for producing
polar molecules with the very low temperatures $T$  ($T \lesssim
1$ mK) and high phase space densities required for most of these
applications.  Production of polar molecules at low translational
temperatures, based on assembly from a sample of laser-cooled
atoms, has recently been demonstrated using photoassociation at $T
\sim 100 \, \mu$K \cite{ourRbCspapers,bagnatoKRb,eylerKRb}, and
may soon be possible using Feshbach resonance-driven
magneto-association at even lower temperatures
\cite{Ingusciomixtures}. Unfortunately, such methods produce
molecular samples that, although translationally and rotationally
cold, have their population in one or several highly-excited
vibrational states. Methods are underway to drive part of the
molecular ensemble to the rovibronic ground state, and to distill
these ground state molecules from the sample.  Such methods may
eventually yield samples with the desired properties. However,
this assembly method is unlikely to become widely useful because
of its expense, complexity, and applicability to a very limited
range of species.

Other groups have developed methods for direct cooling and/or
slowing of molecules
\cite{Doylebuffergastrap,MeijerStarktrap,Rempefilterguide,Chandlerbilliard,YeStarkslower}.
Samples of several molecular species have been produced at $T \sim
10-100$ mK, in all degrees of freedom (translational, vibrational,
rotational), and the resulting molecules have been (or clearly
could be) subsequently trapped.  These direct-cooling methods have
wide applicability and are simpler than the assembly approach.
However, they require a new technological step to reach
temperatures substantially lower than $T \sim 10$ mK. Collisional
cooling of the trapped molecules (sympathetic and/or evaporative)
appears to be the only viable technique for breaking through this
temperature barrier.

Unfortunately, it seems that the prospects are either uncertain,
or demonstrably poor, for further collisional cooling of molecules
in most of the traps now employed.  The essential problem is that
these traps are based on static electromagnetic (EM) fields. Since
static EM fields cannot have a maximum in free space, these traps
hold only weak-field seeking molecules, which are by necessity in
an internally excited state.  Thus, in such traps there is
\textit{always} a channel for inelastic collisions, which can lead
to loss of trapped molecules.  The rotational degree of freedom of
molecules can greatly enhance such inelastic loss mechanisms,
relative to atoms; thus several authors have expressed pessimism
about the prospects for collisional cooling of molecules in
electrostatic traps
\cite{Bohnpolarcollisions,Kajitasemiclassical,Kajitacolder} as
well as magnetostatic traps \cite{Bohnmagnetictrapcollisions}.
While there may be certain situations where the problems with
collisional cooling in static traps can be avoided
\cite{Kajitafermions},  it seems unlikely that such traps could be
used very generally for this purpose.

There have been two types of traps for strong-field seeking
(ground state) molecules discussed previously in the literature.
In the first of these, a centrifugal barrier associated with the
molecules' orbital angular momentum around the trap center (e.g.,
around a charged wire \cite{Starkwiretrap}, or in a storage ring
\cite{Meijerstoragering}) prevents the molecules from colliding
with the electrodes where the strongest electric fields exist.
Such traps are obviously not suitable for collisional cooling to
very low temperatures.  A more promising trap has recently been
demonstrated, based on low frequency modulation of a static trap.
The principle of such traps is similar to that of the Paul trap
for charged particles \cite{RempePaultrap}. However, with
realistic technical parameters, such traps for strong-field
seekers are typically both shallow (with trap depth $D \lesssim
10$ mK) and small (with volume $V \lesssim 10^{-3}$ cm$^3$). These
constraints will make it difficult to load large samples of polar
molecules into such "dipole Paul traps", using currently available
sources of precooled molecules.

In this paper we propose and discuss a new type of trap, based on
the large low-frequency electric polarizability of polar
molecules.  In particular, we argue that it is viable to trap and
collisionally cool molecules in the strong, high-frequency
electric field of a microwave resonant cavity.  Such a microwave
trap has several notable advantages over the traps now employed
for directly-cooled molecules, and may be very useful in the
effort to further cool the samples of polar molecules now
available.  The advantages of the microwave trap include:

\begin{itemize}
 \item ·   The microwave field can trap molecules in their \textit{absolute
 ground state} (a strong-field seeking state), at a free-space maximum
 of the high-frequency electric field.  This in turn eliminates all
 concerns about two-body inelastic collisions leading to loss of
 molecules during evaporative or sympathetic cooling.

 \item ·   With reasonable technical parameters, the microwave trap
 can have large depth $D$ ($D > 1$ K) and volume $V$ ($V > 1$ cm$^3$),
 allowing it to be loaded easily from proven sources of
 directly-cooled molecules \cite{Doylebuffergastrap,MeijerStarktrap,Rempefilterguide,Chandlerbilliard,YeStarkslower}.
 Note that optical dipole traps (which are similar in principle to the
 microwave trap) can also trap ground-state molecules, but optical
 traps are both much shallower and of much smaller volume than the
 microwave trap.

 \item ·   An open trap geometry (using a Fabry-Perot microwave
 cavity) should allow easy overlap of the molecules with laser-cooled
 atoms.  The microwave field will also act as a weak trap for the
 atoms.  Thus sympathetic pre-cooling to temperatures well below
 1 mK seems very promising in the microwave trap.

 \item ·   Strong-field seeking states reside in the region of
 maximum electric field, where they are electrically polarized.
 The resulting dipole-dipole interaction results in huge elastic
 collision cross-sections ($\sigma$), which are desirable for evaporative
 cooling.  This cross-section actually increases as the temperature
 of the molecules decreases ($\sigma \propto T^{-1/2}$) \cite{Kajitasemiclassical}.
 Thus, with sufficiently high initial density, the prospects seem very favorable for bringing
 the trapped molecules to a regime of runaway evaporative cooling.

 \end{itemize}
 In the following sections we will outline the basic principle of
 the microwave trap, in terms of the energy shifts of molecules
 subjected to a microwave field; discuss realistic design parameters
 for an implementation of the trap; and finally, discuss collisions
 within the trap.

\section{Energy level shifts of polar molecules in a
microwave field} \label{sec2}  The basic principle of the
microwave trap is to take advantage of the AC Stark shift
associated with low-frequency transitions arising from rotational
(or inversion-doublet or Lambda/Omega doublet) structure in polar
molecules.  For the purposes of this paper, we confine the
discussion to the effect of the microwave electric field on
diatomic molecules with simple rigid rotor structure (e.g., in
$^1\Sigma$ electronic states), although the ideas can be easily
extended to molecules with more complex structure.  In addition,
we consider primarily the effect on the rotational ground state
($J=0$), although such traps can also be effective for other
states under proper conditions.  We assume the molecule is subject
to a harmonically-varying electric field $\Eb (t)= \textrm{Re}
\{\Eb_0 e^{i \omega t}\} = \mathcal{E}_0 \textrm{Re}\{\epsb e^{i
\omega t}\}$, where $\epsb$ is the complex unit vector indicating
the polarization. Moreover, we assume that trap microwave angular
frequency $\omega$ satisfies $\hbar \omega \lesssim B_e$, where
$B_e$ is the rotational constant.  [$B_e$ is defined such that the
field-free energy $E(J)$ of the state with rotational quantum
number $J$ is $E(J) = B_e J(J+1)$.]  The response of polar
molecules to such low-frequency electric fields is dominated by
the coupling to nearby rotational levels; as such it is a good
approximation to ignore the vibrational and electronic structure
of the molecule. The Hamiltonian of the system is then $H = H_0 +
H'$, where $H_0 = B_e \mathbf{J}^2$, $H' = -\mu \mathbf{n} \cdot
\Eb$, $\mu$ is the electric dipole moment in the molecule-fixed
frame, and $\mathbf{n}$ is the operator indicating the unit vector
along the molecular symmetry axis.  The eigenstates of $H_0$ are
defined by the quantum numbers $J$ and $m = \left\langle J_z
\right\rangle$.

We have performed a general calculation of the AC Stark shift
$\DE$ of the $J=0$ state, which determines the trap depth. Before
describing this general calculation, however, it is instructive to
consider the behavior of $\DE$ in some simple limiting cases. For
low electric field strength ($\mu \mathcal{E}_0 \ll B_e$), $\DE$
can be calculated from standard time-dependent perturbation
theory. To second order,
\begin{eqnarray}
\DE  & = & -\frac{1}{4} \sum\limits_{J',m'}
\frac{\left|{\left\langle{J',m'}\right| H'
\left|{J=0,m=0}\right\rangle}\right|^2}{E(J')-E(0)-\hbar \omega}
\nonumber \\
 & & \; \; \; \; \; \; \; \; \; \; + \; \frac{\left|{\left\langle{J',m'}\right| H'
\left|{J=0,m=0}\right\rangle}\right|^2}{E(J')-E(0)+\hbar \omega}
\nonumber \\
\nonumber \\
 & = & -\frac{\mu^2 \mathcal{E}_2^2}{4} \sum\limits_{m'}
\frac{\left|{\left\langle{1,m'}\right| \mathbf{n} \cdot \epsb
\left|{0,0}\right\rangle}\right|^2}{2B_e -\hbar
\omega} \nonumber \\
 & & \; \; \; \; \; \; \; \; \; \; \; \; \; \; + \; \frac{\left|{\left\langle{1,m'}\right| \mathbf{n} \cdot
\epsb \left|{0,0}\right\rangle}\right|^2}{2B_e +\hbar
\omega} \nonumber \\
 & = & \frac{\mu^2 \mathcal{E}_2^2}{4} \sum\limits_{m'}
\frac{\left|{\left\langle{1,m'}\right| \mathbf{n} \cdot \epsb
\left|{0,0}\right\rangle}\right|^2}{\hbar
\Delta} \nonumber \\
 & & \; \; \; \; \; \; \; \; \; \; \; + \; \frac{\left|{\left\langle{1,m'}\right| \mathbf{n} \cdot
\epsb \left|{0,0}\right\rangle}\right|^2}{\hbar \Delta -2 \hbar
\omega} \label{pertStarkshift}
\end{eqnarray}
where we have used the fact that matrix elements of the form
$\left\langle{J',m'}\right| H' \left|{J,m}\right\rangle$ vanish
unless $J' = J \pm 1$ \cite{TownesandSchawlow},  and we have
defined the detuning $\Delta$ of the microwave field from the $J=0
\leftrightarrow J=1$ resonance: $\Delta \equiv \omega - 2B_e$.
This perturbative expression makes clear some of the salient
features of the microwave trap. In particular, it is evident that
for red detuning ($\Delta < 0$), $\DE < 0$; thus in this case $J =
0$ molecules are attracted to regions of strong microwave electric
field. Moreover, $\DE$ (which will be the trap depth) can be
enhanced by operating at small detuning.

The microwave trap is clearly similar to the more-familiar optical
dipole trap, which is used commonly for laser-cooled atoms
\cite{opticaldipoletraps}.   For optical traps, it is impractical
to operate in the regime of small detuning $\hbar
\left|{\Delta}\right| \lesssim \mu \mathcal{E}_0$, since in this
case a high rate of spontaneous emission from the nearby,
short-lived electronically excited state leads to rapid heating of
the sample. The expression for the depth of an optical dipole trap
is thus typically written in the perturbative limit as above. By
contrast, the spontaneous-emission lifetimes of rotational states
are much longer than typical trap lifetimes; thus heating by
photon scattering cannot be troublesome in microwave traps. This
makes it possible, in principle, to operate at arbitrarily small
detunings, where trap depth is maximized.

     It is also instructive to consider the limit of low frequency,
$\hbar \omega \ll B_e$.  Here, for a linearly-polarized field
$\epsb = \hat{z}$, $\DE$ must be the same as the DC Stark shift
$\DE_{DC}$ in an applied DC electric field $\Eb_{DC} = \left(
{\mathcal{E}_0 / \sqrt{2}} \right) \hat{z}$. (The factor of
$1/\sqrt{2}$ arises from the fact that the molecule responds, on
average, to the r.m.s. microwave field strength; this is in turn
because the Stark effect is a second-order effect in the low-field
limit.) Indeed, Eq.(\ref{pertStarkshift}) reproduces the familiar
perturbative expression for $\DE_{DC}$ in the limit $\omega
\rightarrow 0$ \cite{TownesandSchawlow}. A numerical calculation
of DC Stark shifts in larger fields ($\mu \mathcal{E}_{DC} \gtrsim
B_e$) is straightforward, and has been presented in, e.g., Ref.
\cite{DeMilleQC}. This calculation yields the result that
$\DE_{DC} \sim -(\mu /2)\mathcal{E}_{DC}$ for large electric
fields. This is in accord with a simple physical picture in which
the large field almost completely polarizes the $J=0$ state, which
then responds in the same way as a fixed electric dipole in an
external electric field. From this picture, it can be anticipated
that even for higher frequencies $\hbar \omega \sim B_e$,
microwave traps should have similar shifts, $\DE \sim -(\mu
/2)\mathcal{E}_0$, for moderately strong fields $\mu \mathcal{E}_0
\gtrsim B_e$.

Neither of the limiting cases discussed so far is sufficient to
describe the optimal conditions for the microwave trap.  In order
to obtain large trap depths it is necessary to have $\mu
\mathcal{E}_0 \gtrsim B_e$, so that the low-field limit does not
apply. In addition, the desire to trap the molecules in a
free-space maximum of $\mathcal{E}_0$ means that the dimensions of
the trap must be at least as large as the microwave wavelength
$\lambda = 2 \pi c/ \omega$. For practical trap sizes, this
requires $\hbar \omega \sim B_e$; this is preferable in any case,
to maximize $\DE$ by operating at small detuning $\Delta$.

\begin{figure*}
\includegraphics[width=18.0cm] {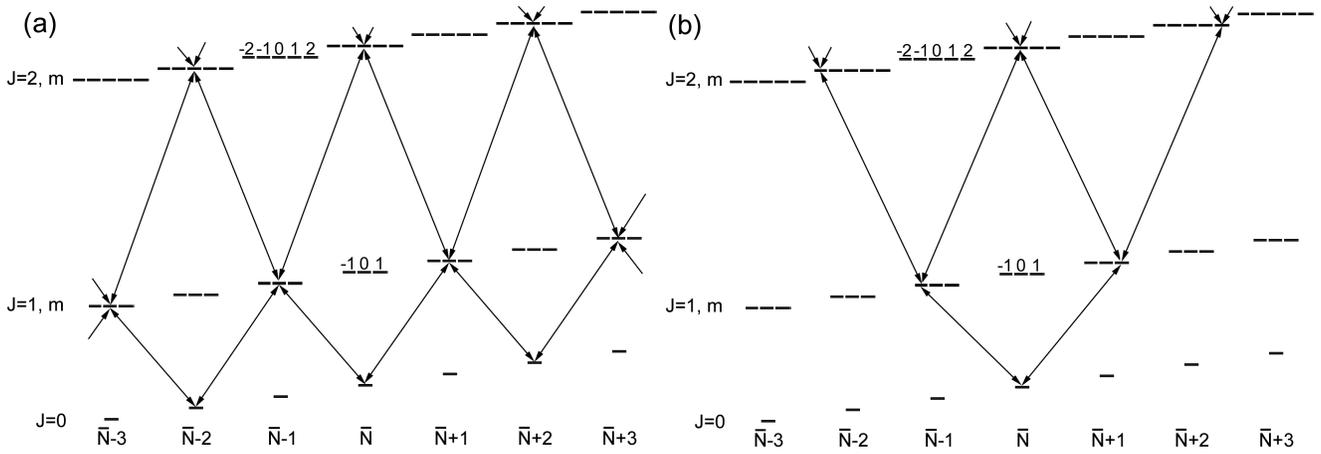}
\caption{Dressed-state energy level diagrams for rotational states
of a diatomic molecule subject to a microwave electric field.  The
indices at the bottom indicate the photon number $\bar{N} + n$,
where $\bar{N}$ is the average number of photons in the field; the
labels $J, m$ at the side indicate the rotational angular momentum
and its $z$-projection, respectively.  Arrows indicate
$1^{st}$-order couplings between levels due to $H'_q$;
unterminated arrows couple to levels outside the range of the
figure.  The subset of the Hilbert space used for the calculations
described in the main text consists of only levels connected to
$\left| {\psi_{000}} \right\rangle = \left| {0,0} \right\rangle
\left| {\bar{N}} \right\rangle$ with some number of arrows. For
clarity, we have chosen $\hbar \omega = 0.2 B_e$ for this figure.
(a) Linear polarization $\epsb = \hat{z}$. (b) Circular
polarization $\epsb = (\hat{x}-i\hat{y})/\sqrt{2}$. Note that the
linearly-polarized case contains many more states, and inevitably
includes near-degeneracies at some values of $n$.} \label{figure1}
\end{figure*}

We have calculated $\DE$ for a more general range of $\omega$ and
$\mathcal{E}_0$, using the dressed-state formalism
\cite{CohenTannoudji}.   The classical-field Hamiltonian $H$ is
replaced with its quantized-field analogue: $H_q = H_{0q} + H'_q$,
where
\begin{equation}
 H_{0q} = B_e \mathbf{J}^2 + \hbar \omega
(\hat{N}-\bar{N})
\end{equation}
and
\begin{equation}
H'_q = -\mu \mathbf{n} \cdot \Eb = -\mu \sqrt{\frac{\hbar
\omega}{2 \omega}{2 \epsilon_0 V}}\left[ {(\mathbf{n} \cdot
\epsb^*) a^{\dag} + (\mathbf{n} \cdot \epsb) a} \right].
\end{equation}
Here we have introduced the photon creation and destruction
operators $a^{\dag}, a$, and the photon number operator $\hat{N} =
a^{\dag}a$.  (The parameter $V$ refers to the volume of a
fictitious box used to define the boundary conditions of the
quantized electromagnetic field; as usual in such calculations,
all relevant physical quantities are independent of $V$.) The
energy scale is defined relative to the mean number of photons in
the field, $\bar{N} = \frac{\mathcal{E}_0^2}{2} \frac{\epsilon_0
V}{\hbar \omega}$. Throughout we are interested in strong fields,
so that $\bar{N} \gg 1$.  We use as basis states the
molecule+field eigenstates of $H_{0q}$: $\left| {\psi_{Jmn}}
\right\rangle = \left| {J,m} \right\rangle \left| {\bar{N} + n}
\right\rangle$, where $\bar{N} + n$ is the dressed-state photon
number. The diagonal matrix elements of $H_q$ are $\left\langle
{\psi_{Jmn}} \right| H_q \left| {\psi_{Jmn}} \right\rangle = B_e
J(J+1) + n \hbar \omega$. Off-diagonal matrix elements of the form
$\left\langle {\psi_{J'm'n'}} \right| H_q \left| {\psi_{Jmn}}
\right\rangle$ vanish unless $J' = J \pm 1$ and $n' = n \pm 1$.
Explicitly, using the approximation that $\sqrt{\bar{N} \pm n}
\approx \sqrt{\bar{N} \pm n \pm 1} \approx \sqrt{\bar{N}}$, we can
write
\begin{eqnarray}
\left\langle {\psi_{J'm'n'}} \right| H_q \left| {\psi_{Jmn}}
\right\rangle = \nonumber \\
(\mu \mathcal{E}_0 / 2) \left\{ \left\langle {J',m'} \right|
\mathbf{n} \cdot \epsb \left| {J,m}
\right\rangle \delta_{n', n-1}   + \right.  \nonumber \\
\left. \; \; \; \; \left\langle {J',m'} \right| \mathbf{n} \cdot
\epsb^* \left| {J,m} \right\rangle \delta_{n', n+1} \right\}.
\label{dressedstatematrixelement}
\end{eqnarray}
The relevant quantity $\DE$ is the shift of any state as a
function of $\mathcal{E}_0$; this should be the same for any value
of $|n|$ such that $|n| \ll \bar{N}$, so for convenience we focus
on small values of $n$. For numerical calculations, the Hilbert
space is truncated to a finite range $0 \leq J \leq J_{max}$ and
$-n_{max} \leq n \leq n_{max}$; the resulting Hamiltonian matrix
is numerically diagonalized using standard linear algebra
routines. Convergence is verified in two ways: by checking that
the shifts of states $\left| {J,m} \right\rangle \left| {\bar{N}}
\right\rangle $ and $\left| {J,m} \right\rangle \left| {\bar{N}
\pm 2} \right\rangle$ are the same, and by checking that the
results are not affected by substantially expanding the Hilbert
space.

We first consider the case of a linearly polarized field, $\epsb =
\hat{z}$. In this case the molecular part of the off-diagonal
matrix element is given by
\begin{eqnarray}
\left\langle {J',m'} \right| \mathbf{n} \cdot \epsb \left| {J,m}
\right\rangle  = \left\langle {J',m'} \right| \mathbf{n} \cdot
\epsb^* \left| {J,m} \right\rangle  = \nonumber \\
\left\langle {J',m'} \right| \cos{\theta} \left| {J,m}
\right\rangle = \int{Y_{J'}^{m'^*} \cos{\theta} \, Y_J^m d\Omega}
\nonumber = \\ \left\{
   \frac{J}{\sqrt{2J-1}\sqrt{2J+1}} \, \delta_{J',J-1} + \right. \nonumber \\
   \left. \frac{J}{\sqrt{2J+1}\sqrt{2J+3}} \, \delta_{J',J+1} \right\}
\delta_{m',m}. \label{linmatrixelt}
\end{eqnarray}
Thus only states with $m = 0$ couple at any order to the states
$\left| {\psi_{00n}} \right\rangle$  of interest; Fig.
\ref{figure1}(a) shows the manifold of coupled states that
comprise the relevant Hilbert space in this case.

\begin{figure*}
\includegraphics[width=18.0cm] {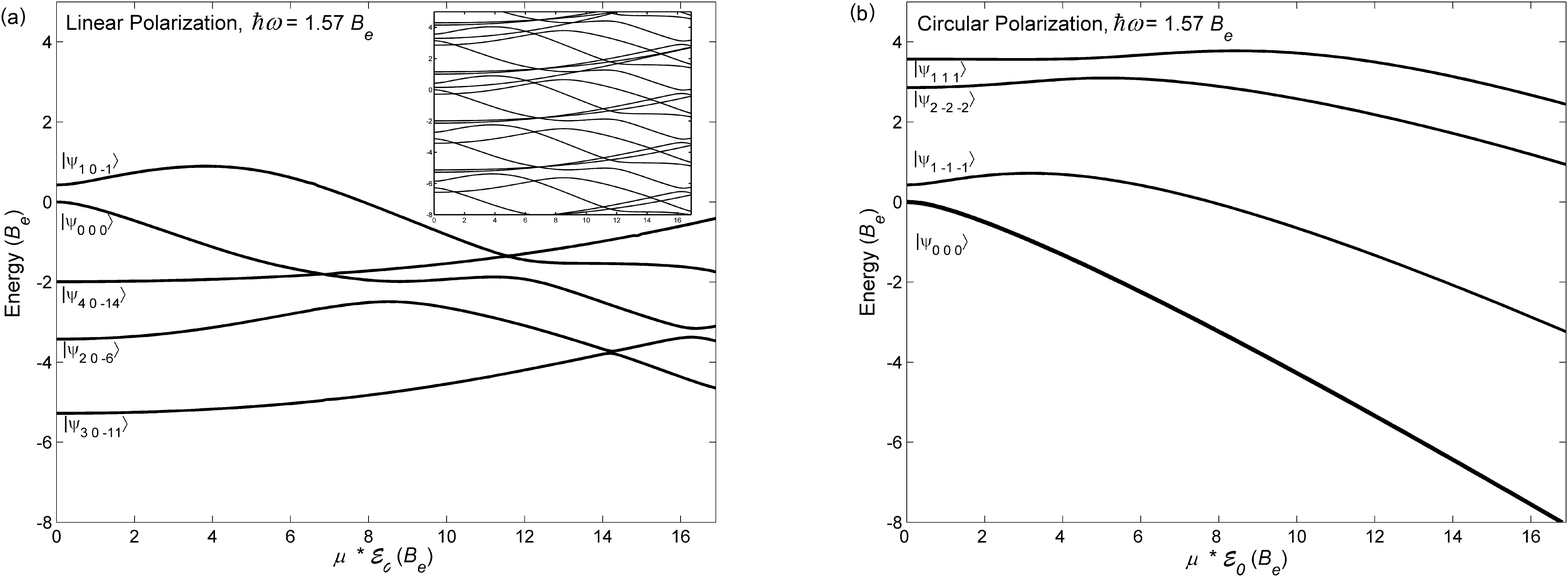}
\caption{Energies of dressed states vs. applied microwave electric
field strength.  State labels are zero-field basis states $\left|
{\psi_{Jmn}} \right\rangle$, as discussed in the main text.  Only
the subspace connected to $\left| {\psi_{000}} \right\rangle$ is
shown.  For this figure we have chosen a typical working value of
the microwave frequency, $\hbar \omega = 1.57 B_e$.  (a) Linear
polarization $\epsb = \hat{z}$. In this calculation, $J_{max} = 5$
and $n_{max} = 20$.  For clarity, only one dressed state for each
initial value of $J < J_{max}$ is shown in the main diagram. (The
inset shows the full Hilbert space included in the calculation;
the pattern of levels is repeated for states with photon number
$n'=n \pm 2k$, where $k$ is any integer.) Note the large avoided
crossings with the state that originates as $\left| {\psi_{000}}
\right\rangle$ at zero field; the crossing of states labeled
$\left| {\psi_{000}} \right\rangle$ and $\left| {\psi_{4 \, 0 \,
-14}} \right\rangle$ is also avoided, but the separation is too
small to see in this diagram. (b) Circular polarization $\epsb =
(\hat{x}-i\hat{y})/\sqrt{2}$. In this calculation, $J_{max} = 10$
and $n_{max} = 10$. All states of the coupled subspace in the
plotted energy region are shown.  Note the monotonic behavior and
lack of avoided crossings for the $\left| {\psi_{000}}
\right\rangle$ state.} \label{figure2}
\end{figure*}

In the low-frequency limit $( \hbar \omega \ll B_e$ and $\hbar
\omega \ll \mu \mathcal{E}_0 )$, our calculation reproduces the
numerical DC Stark shift calculations, as expected, for both weak
and strong fields. At higher frequencies, more complex behavior is
observed: multiple avoided crossings, of widely varying width,
occur between states of (nominally) different values of $J$ and
$n$.  A typical case is shown in Fig. \ref{figure2}(a). This
behavior can be understood easily in the classical-field picture.
In analogy to the DC field case, as $\mathcal{E}_0$ increases from
zero, the $\left| { J=0,m=0 } \right\rangle$ state decreases in
energy, while all other $\left| { J',m'=0 } \right\rangle$ states
increase.  It is thus inevitable that, as $\mathcal{E}_0$
increases, the condition will be met for an exact k-photon
resonance between $\left| { J=0,m=0 } \right\rangle$ and any other
state $\left| { J',m'=0 } \right\rangle$.  It is this resonance
which gives rise to the avoided crossing between dressed states
$\left| {\psi_{00n}} \right\rangle$ and $\left| {\psi_{J' 0
(n-k)}} \right\rangle$ .  If the resonance occurs at large enough
values of $\mathcal{E}_0$ (such that $\mu \mathcal{E}_0 \sim J'
B_e$), the effect of the anticrossing is comparable in size to the
overall shift from zero field ($\sim \mu \mathcal{E}_0$).

The presence of these avoided crossings makes microwave traps
using linearly polarized fields rather unattractive.  Physically,
as a molecule moves from regions of low to high field strength
(either spatially or temporally), it is susceptible to absorption
of multiple microwave photons, resulting in both a significant
degree of rotational excitation, and a reduction in effective trap
depth relative to our previous expectations.  This situation
corresponds to adiabatic following of the energy curve through an
avoided crossing, leading to effective transfer from the initial
state $\left| {\psi_{00n}} \right\rangle$ to a final state $\left|
{\psi_{J' 0 (n-k)}} \right\rangle$. From simple estimates of the
time scales for change of $\mathcal{E}_0$ under likely physical
conditions, we find that adiabatic following is probable even for
crossings that are too small to observe in Fig. \ref{figure2}(a),
so that a reliable prediction for the behavior of the system may
be difficult without very detailed modeling of the molecular
trajectories.  We note in passing that this conclusion regarding
linearly polarized fields may be too pessimistic for molecules
with different structure, such as an inversion doublet with
splitting much less than the rotational constant (where the
microwave frequency should be detuned to the red of the doublet
transition).  However, we have not tried to analyze such a case
explicitly.

Remarkably, the desired simple behavior of the microwave trap can
be recovered by using a circularly polarized field, e.g. with
$\epsb = (\hat{x} - i \hat{y})/ \sqrt{2}$. In this case,
absorption (emission) of k photons is necessarily accompanied by a
change in angular momentum projection $\Delta m = -k (k)$.  The
critical difference from the case of linear polarization is that
for the $\left| { J=0,m=0 } \right\rangle$ state of interest, net
absorption or emission of $k$ photons requires coupling to a state
with angular momentum $J' \geq m' = k$, which is, for any field
strength, separated in energy from the $\left| { J=0,m=0 }
\right\rangle$ state by $E(J')-E(J=0)
> 2B_e k$.  Thus, for a red detuned microwave trap
(with $\hbar \omega < 2B_e$), there are simply no resonant
multiphoton transitions possible from the $\left| { J=0,m=0 }
\right\rangle$ state.  This dramatically simplifies the behavior
of $\DE$ as a function of $\mathcal{E}_0$, compared to the case of
linear polarization.  The manifold of coupled states that comprise
the relevant Hilbert space for the case of circular polarization
is shown in Fig. \ref{figure1}(b).

The molecular part of the off-diagonal matrix elements used for
explicit calculations of the circular-polarization case are
\begin{eqnarray}
\left\langle {J',m'} \right| \mathbf{n} \cdot \epsb^* \left| {J,m}
\right\rangle = \left\langle {J,m} \right| \mathbf{n} \cdot \epsb
\left| {J',m'} \right\rangle = \nonumber \\
\frac{1}{\sqrt{2}} \left\langle {J',m'} \right| \sin{\theta} e^{i
\varphi} \left| {J,m} \right\rangle = \nonumber \\
  \frac{1}{\sqrt{2}} \int{Y_{J'}^{m'^*} \sin{\theta}
e^{i \varphi} Y_J^m d\Omega} = \nonumber \\
\left\{
   \frac{\sqrt{J-m-1} \sqrt{J-m}}{\sqrt{2J-1}\sqrt{2J+1}} \, \delta_{J',J-1} + \right. \nonumber \\
   \left. \frac{\sqrt{J+m+1} \sqrt{J+m+2}}{\sqrt{2J+1}\sqrt{2J+3}} \, \delta_{J',J+1} \right\}
\delta_{m',m+1}.
\end{eqnarray}
Typical results of the calculation are shown in Fig.
\ref{figure2}(b). For moderately large fields such that $\mu
\mathcal{E}_0 \gtrsim B_e$, we find the desired (and originally
expected) behavior $\DE \approx -\alpha \mathcal{E}_0$, where the
proportionality factor $\alpha$ is roughly constant over a wide
range of $\mathcal{E}_0$, is of order $\alpha \sim \mu /2$, and is
maximal when the detuning $\Delta$ is minimized.

\section{Realistic design parameters for a microwave trap}
\label{sec3} The requirement for a circularly-polarized field,
along with a desire to maintain an open geometry for optical and
other access to the trap region, has led us to consider
Fabry-Perot type resonators for the microwave trap.  Such
resonators are common in the mm-wave \cite{microwaveFPreview}
through optical regions \cite{Yariv}, but rather less so in the
microwave regime. Nevertheless, cavities with characteristics very
similar to those required have been demonstrated
\cite{GrachevFP,gridinputcoupler}; we closely follow the treatment
of Ref. \cite{GrachevFP} in our discussion here. Our goal is to
outline a basic, realistic design for such a resonator, and
discuss the volume, depth, and other characteristics of the
resulting trap potential.

\begin{figure*}
\includegraphics[width=18.0cm] {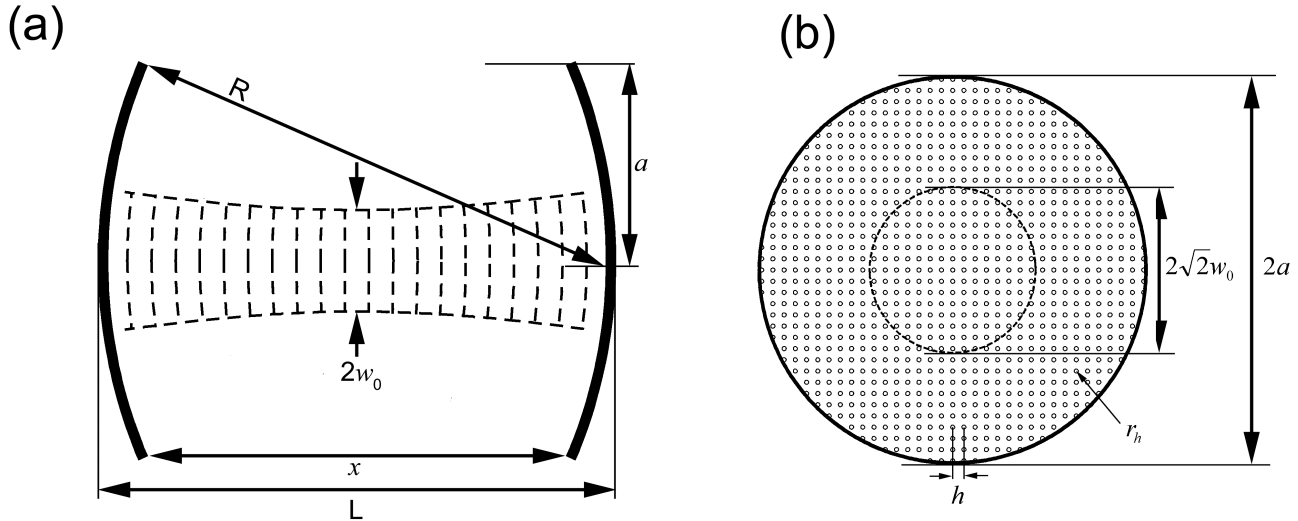}
\caption{Layout of the microwave Fabry-Perot cavity.  Labeled
dimensions are defined in the main text.  The figures are in scale
with the nominal dimensions described in the text.  (a) View from
side.  The dashed lines indicate the TEM$_{00}$ field mode.  (b)
View from input end.  The grid of holes for partial transmission
of the incident power is shown.  The beam spot size at the mirror
is $2w(z) = 2\sqrt{2}w_0.$} \label{figure3}
\end{figure*}

We consider a symmetric, spherical-mirror Fabry-Perot cavity.  We
assume a confocal geometry, where the mirrors have spacing $L$ and
radius of curvature $R = L$, and define the $z$-axis along the
symmetry axis of the resonator, with $z = 0$ midway between the
mirrors.  We consider the lowest-order transverse mode
(TEM$_{00}$)
 in this cavity, for which the resonant condition is $L =
(q+1/2)\lambda /2$, where $q$ is the integer number of
half-wavelengths along the cavity axis; we choose $q$ odd to
produce an antinode of the microwave electric field at $z = 0$.
For $q \gg 1$ so that the paraxial-ray approximation holds
\cite{paraxialrayrefs}, the electric field magnitude at time $t$
and position $(z,\rho ,\phi)$ in cylindrical coordinates can be
written as
\begin{eqnarray}
\Eb (z, \rho , \phi) = \mathcal{E}_{00}
\frac{w_0}{w(z)} e^{-\rho^2/w^2(z)} \cdot \nonumber \\
\cos{ \left( kz - \textrm{tan}^{-1}(z/z_0) + \frac{k \rho^2}{2
\mathcal{R}(z)} \right) } \textrm{Re}\{ \epsb e^{-i \omega t}\}.
\label{gaussianbeam}
\end{eqnarray}
Here we have introduced the wavevector $k = 2 \pi / \lambda$ and
the standard Gaussian beam parameters: the minimum beam spot size
$w_0$ and general spot size $w(z)=w_0 \sqrt{1 + z^2/z_0^2}$; the
wavefront radius of curvature $\mathcal{R}(z) = z(1+z_0^2/z^2)$;
and the Rayleigh range $z_0 = \pi w_0^2/\lambda$. For a confocal
cavity, $w_0 = \sqrt{\lambda L / (2 \pi)} = \sqrt{Lc/\omega}$, and
$z_0 = L/2$. In the region of interest around the center of the
resonator, the magnitude of the electric field is
well-approximated by the expression
\begin{eqnarray}
\mathcal{E}_0(z,\rho) = \mathcal{E}_{00}e^{-\rho^2/w_0^2}
\cos{(2\pi
z/\lambda)} \approx \nonumber \\
\mathcal{E}_{00}(1 - \frac{1}{2} \frac{8\pi}{(q+1/2) \lambda^2}
\rho^2 - \frac{1}{2} \frac{4 \pi^2}{\lambda^2} z^2).
\label{efieldintrapcenter}
\end{eqnarray}
Thus, the trap formed by the field antinode around the origin is
approximately harmonic in all directions, and has volume of $\sim
\lambda^3$. Fig. \ref{figure3} shows the geometric layout of the
trap.

Suppose the input and output mirrors have amplitude transmission
coefficients $t_{in} = t$ and $t_{out} = 0$, respectively; loss
coefficients $\gamma_{in} = \gamma_{out} = \gamma \ll 1$; and
reflection coefficients $r_{in} = \sqrt{1 - \gamma^2 - t^2}$ and
$r_{out} = r \sqrt{1 - \gamma^2}$, respectively. For mirrors
constructed of metal with resistivity $\rho$, it is easily shown
that $\gamma = (8 \epsilon_0 \rho \omega)^{1/4}$ \cite{Jackson}.
The nonzero transmission $t$ of the input mirror can be engineered
by perforating the mirror with an array of small holes of radius
$r_h \ll \lambda /4 \pi$, in a square grid with spacing $h <
\lambda$. This yields a transmission coefficient $t = \frac{16
\pi}{3} \frac{r_h^3}{h^2 \lambda}$ \cite{GrachevFP}. These
conditions require that the thickness of the metal, $d$, satisfy
$d \ll \delta \ll 2r_h$, where $\delta$ is the skin depth of the
metal at frequency $\omega$: $\delta = \sqrt{2 \rho / (\mu_0
\omega)}$. We assume that the mirrors are of sufficiently large
radius $a$ that diffractive losses $\gamma_d$ are negligible; this
is satisfied when $\gamma_d = e^{-a^2/(2 w_0^2)} \ll \gamma$.

It is straightforward to show that, for a given microwave power
incident on the input mirror of such a cavity, the electric field
in the cavity is maximized when $t^2 = 2 \gamma^2$; in this case
the cavity has loaded $Q$-factor $Q_l = \pi (q+1/2)/(2 \gamma^2)$;
the unloaded $Q$-factor, for the cavity without the input coupling
grid, is $Q_0 = 2Q_l$.  We assume microwave power $P_{in}$ is
incident on the cavity input mirror, in the form of a mode-matched
Gaussian beam with electric field distribution
$\mathcal{E}_{in}(\rho, z=L/2)=\left( {\mathcal{E}/\sqrt{2}}
\right) e^{-\rho^2/(2 w_0^2)}$; here $P_{in} = \pi w_0^2
\sqrt{\frac{\epsilon_0}{\mu_0}}\frac{\mathcal{E}_i^2}{4}$. In this
case, the electric field at the center of the cavity is given by
$\mathcal{E}_0 = \sqrt{2} \mathcal{E}_i / \gamma$.

Next, in order to convey a sense of realistic parameters for the
trap, we choose a specific set of nominal values for all the
relevant parameters.  We envision a microwave field at frequency
$\omega = 2 \pi \times 15$ GHz, with wavelength $\lambda = 2$ cm.
(This is a convenient range for many oxide, fluoride, and nitride
species.) For $q =$ 21 half-wavelengths in the cavity, $L = 21.5$
cm and the beam waist size at the center (end) of the cavity is
$w_0 =$ 2.61 cm [$w(z=L/2) =$ 3.70 cm].  For room-temperature
copper mirrors, using $\rho_{Cu} = 1.7 \times 10^{-8} \Omega
\cdot$m, we find $\gamma = 1.8 \times 10^{-2}$ and $Q_l = 1.0
\times 10^5$. We take the mirror radius $a = 0.4L = 8.6$ cm; this
gives $\gamma_d = 4.4 \times 10^{-3} \ll \gamma$. The required
input coupling can be achieved with $h = 0.5$ cm, $r_h = 0.092$
cm.  We believe that fabrication of a cavity with such parameters
will be straightforward.

We assume a microwave input power $P_{in} = 2$ kW.  Such levels of
power are commonly available from klystron-based satellite
communication amplifiers, throughout much of the microwave band,
i.e., in the frequency range of roughly 2-18 GHz.  (Available
power is a factor of 5-10 lower for higher frequencies up to $\sim
50$ GHz.)  These amplifiers typically deliver the microwave power
into standard waveguide; the required Gaussian mode for input to
the cavity can be effectively obtained by launching into a
corrugated scalar feed horn \cite{scalarfeedhorn} or, less
efficiency, with a standard pyramidal horn.  Assuming perfect
mode-matching at the input, we find a value for the peak electric
field inside the cavity of $\mathcal{E}_0 = \mathcal{E}_{00} =
2.8$ MV/m.  The parameter $\mu \mathcal{E}_0$, which governs the
trap depth for a molecule with dipole moment $\mu$, is then $\mu
\mathcal{E}_0 = 0.48$ cm$^{-1} \mu$[D] = 0.69 K $\mu$[D].  Thus,
for a typical range of parameters (such that $\mu \mathcal{E}_0
\gtrsim B_e$ and thus $\DE \sim -\mu\mathcal{E}_0/2$), such a
microwave trap can achieve trap depths $\DE \gtrsim 1$ K for
molecular species with dipole moments $\mu \gtrsim 4$ D. We are
currently constructing experiments optimized to trap the highly
polar species SrO, with $\mu = 8.9$ D and $B_e = 0.34$ cm$^{-1}$
\cite{Herzberg}. In this case we find explicitly a trap depth $\DE
= -2.8$ K.

We have envisaged two possibilities for loading such a trap.  In
the simplest scheme, the trap is left off until a pulse of cold
molecules enters the trapping region, at which time the microwave
field is rapidly turned on.  Note that the time scale to build up
energy in the trap is $\tau = Q_l/\omega$; for our nominal
parameters, $\tau = 2 \; \mu$s. This is much shorter than the
typical time for a cold molecule to traverse the trapping region,
so molecules can be effectively stopped when the trap potential
suddenly appears. This "trap door" loading scheme requires a
source that delivers pulses of cold molecules already in their
strong-field seeking ground state, such as the alternating
gradient decelerator \cite{altgrad}, or a source of weak-field
seekers followed by a coherent pulse of resonant microwaves to
transfer population to the ground state.

In the trap door scheme, it is not usually profitable to load more
than a single pulse of molecules into the trap, since
already-trapped molecules can escape while the trap is turned off
for loading of subsequent pulses.  However, it may be possible to
load multiple pulses of weak-field seeking molecules into the
microwave trap.  Here, we envision optically pumping the molecules
from the weak-field seeking excited state in which they are
delivered, and into the trapped ground state
\cite{optpumploading}.  The dissipation associated with optical
pumping, combined with the rather different trap potentials for
the strong- and weak-field seekers, makes the loading of one pulse
completely transparent to any existing trapped molecules.  Note
that, in the absence of applied fields, parity selection rules
forbid optical pumping from $J=1$ levels to the $J=0$ ground
state.  However, the presence of the microwave field breaks this
selection rule, and makes it possible to perform the desired
pumping.  Such a scheme makes it attractive to load the trap using
sources of weak-field seekers, such as the Stark decelerator
\cite{originalStarkslower} or the quadrupole filter + guide
\cite{Rempefilterguide}. For our experiments with SrO, we plan to
use a version of the quadrupole guide, using a buffer-gas
precooled source of laser-ablated SrO \cite{Doylecoldbeam}.

\section{Collisions in the microwave trap}
\label{sec4} As noted in the introduction, one of the main
advantages of the microwave trap is that it holds molecules in
their absolute ground state, so that two-body inelastic collisions
are impossible.  Here we point out that, in addition, molecules
held in a microwave trap may be subject to very high rates of
elastic collisions.  This is attractive, since the
re-thermalization associated with elastic collisions is a key step
in the process of evaporative and/or sympathetic cooling.

The primary point of the discussion is that, in the center of the
microwave trap, the molecules are subject to the large microwave
electric field, which results in their nearly-complete electrical
polarization.  The r.m.s. expectation value of the dipole moment
of the molecular state in the presence of the microwave field,
$\left\langle {\mu} \right\rangle$, can be obtained simply from
the earlier calculations; it is given simply by the slope of the
ground-state energy vs. electric field curve, $\left\langle {\mu}
\right\rangle =
\partial(\DE)/\partial \mathcal{E}_{rms} = \sqrt{2} \partial(\DE)/\partial \mathcal{E}_0 $.
Under typical conditions, throughout most of the volume of the
trap, $\left\langle {\mu} \right\rangle \sim \mu/2$. The bare
dipole-dipole interactions between the polarized molecules are
both very strong and of long range; this in turn greatly enhances
the elastic collision cross-sections, compared to the case between
unpolarized molecules.

Kajita \cite{Kajitasemiclassical,Kajitacolder}, and Bohn
\cite{Bohnpolarcollisions,BohnOHcollisions}, have considered
elastic collisions between electrically polarized molecules in a
variety of different regimes, using rather different techniques.
We are particularly interested here in collisions between
strongly-polarized molecules at rather high temperatures ($T \gg
1$ mK, corresponding to the initial conditions in the trap).  In
this regime, Kajita has argued that a semi-classical calculation
is valid \cite{Kajitasemiclassical}.  Although the discussion of
Ref. \cite{Kajitasemiclassical} explicitly focuses on collisions
between weak-field seeking states in a static electric trap, it is
straightforward to recast the elastic cross-section derived there
in terms of the parameters used in this paper.  Specifically,
using from Ref. \cite{Kajitasemiclassical} Eqs.(7), (8), and (10),
Table 1, and the discussion following Eq.(10), we find the
dipole-dipole elastic collision cross-section $\sigma$ between
molecules of relative velocity $v$ and mass $m$ can be written as
\begin{eqnarray}
\sigma & \approx & \frac{40 \pi \sqrt{2}}{3} \frac{1}{4 \pi
\epsilon_0} \frac{\left\langle {\mu} \right\rangle^2}{\hbar v}
\nonumber \\ & \approx & \sqrt{\frac{m \textrm{[amu]}}{T
\textrm{[K]}}} \left\langle {\mu \textrm{[D]}} \right\rangle^2
\cdot 4 \times 10^{-12} \textrm{cm}^2. \label{crosssection}
\end{eqnarray}
For the specific case of SrO, where $\left\langle {\mu}
\right\rangle \approx 6.9$ D, $m = 104$ amu, and (when initially
trapped) $T = \DE = 2.8$ K, this yields a remarkably large
cross-section $\sigma \approx 1.2 \times 10^{-9}$ cm$^2$. This is
an extraordinarily large cross-section, which will lead to a
substantial elastic collision rate even with a relatively small
number of initially-loaded molecules. We stress again the notable
fact that $\sigma \propto T^{-1/2}$; this means that, as the
sample cools, collision rates can be preserved with little
difficulty.

The regime in which the result of Eq.(\ref{crosssection}) is valid
appears to be rather broad.  In particular, the semiclassical
treatment (which assumes contributions from many partial waves)
should break down significantly only when $\sigma \lesssim
\sigma_q$, where $\sigma_q = 4\pi(\hbar/mv)^2$ is the maximum
quantum cross-section for a single partial wave. Using the
numerical relation of Eq.(\ref{crosssection}), we find that the
condition $\sigma / \sigma_q > 1$ for the presumed validity of the
semiclassical calculation implies a very weak condition on the
temperature: $T > (m\textrm{[amu]})^{-3} \left\langle
{\mu\textrm{[D]}} \right\rangle^{-4} \cdot 64 \; \mu$K. Returning
again to the example of SrO, $\sigma / \sigma_q
> 1$ for $T \gtrsim 3 \times 10^{-14}$ K! It appears that the
semiclassical relation of Eq.(\ref{crosssection}) is likely to
remain valid under most realistic experimental conditions.  It is
notable that the scaling with $T$ appears to be borne out in the
explicit calculations of Ref. \cite{BohnOHcollisions}, for
temperatures down to $T \sim 1 \; \mu$K.

We point out, parenthetically, that the result of
Eq.(\ref{crosssection}) can be approximately derived using an
extremely simple model.  We anticipate that in the semiclassical
limit, the anisotropy of the dipole-dipole interaction will play a
minor role in the average scattering properties.  This leads us to
consider the cross-section for scattering by an isotropic
potential of the form $V_{iso}(r) =
-\left\langle{\mu}\right\rangle^2/(4\pi\epsilon_0 r^3)$. It was
shown by Julienne and Mies that for such a potential, a
semiclassical description of scattering should be valid for sample
temperatures
\begin{equation}
T > T_Q = \frac{2^{18}}{3^{12}} \frac{\hbar^6 (4 \pi
\epsilon_0)^2} {m^{3} \left\langle{\mu}\right\rangle^{4}};
\label{JulienneT_Q}
\end{equation}
for SrO under our conditions, $T_Q \approx 2 \times 10^{-13}$ K,
in reasonable agreement with the estimate above for the regime in
which the semiclassical description is valid. In the extreme
high-temperature limit ($T \gg T_Q$), the classical path of the
scattered particle should be close to a straight line; this makes
it possible to use the Eikonal approximation \cite{Sakurai}, where
(by invoking the optical theorem), the total elastic scattering
cross-section can be written as
\begin{equation}
\sigma_{Eik} = -4\pi\int\limits_{0}^{\infty}{b(\cos{(2\Delta(b))}
- 1)db}, \label{Eikcrosssection}
\end{equation}
with \begin{equation} \Delta(b) = -\frac{1}{2v}
\int\limits_{-\infty}^{\infty}{V_{iso}\left({\sqrt{b^2+z^2}}\right)
dz}. \label{Eikdef}
\end{equation}
This yields the simple analytic solution
\begin{equation}
\sigma_{Eik} = 2 \pi^2 \left\langle {\mu} \right\rangle^2 / (4 \pi
\epsilon_0 \hbar v). \label{Eiksolution}
\end{equation}
 The Eikonal solution for the anisotropic
potential exhibits the same scaling as the proper semiclassical
solution, and is numerically smaller by a factor of only $\sim 3$.
It is not clear whether this difference arises principally because
we have neglected the anisotropy of the interaction (particularly
the repulsive part of the potential), or because the classical
path deviates significantly from a straight line. Nevertheless, we
find this derivation useful, for the straightforward way it leads
to the correct scaling (and nearly correct magnitude) of the
cross-section.

Finally, we point out the suitability of the microwave trap for
performing sympathetic cooling of the trapped molecules, by
contact with much colder laser-cooled atoms.  The geometry of our
proposed trap is very convenient for this purpose.  The open
distance between the edges of the mirrors (see Fig. 3) is $x = 2L
\left({\sqrt{1-(a/L)^2} - 1/2}\right) = 17.9$ cm , which is
greater than the mirror diameter ($2a = 17.2$ cm). Thus, there is
sufficient open area to overlap 3 orthogonal laser beams in the
center of the microwave trap, as required for a superposed atomic
magneto-optic trap (MOT).  In addition, the microwave field itself
will form a weak, conservative trap for the atoms, analogous to
that formed by a far off-resonant optical dipole trap.  The
microwave polarizability of atoms is virtually identical to the DC
polarizability; the Stark shift of the atomic ground state in the
presence of a DC electric field $\mathcal{E}_{DC}$  is typically
written in the form $\DE_{atom} = -\alpha_0 \mathcal{E}_{DC}^2 /
2$; for atomic Cs, $\alpha_0 = 6.61 \times 10^{-39}$ C $\cdot$
m$^2$/V \cite{Cspol}. With our nominal trap parameters, and taking
into account that the atoms respond to the r.m.s. AC electric
field, $\mathcal{E}_{rms} = \mathcal{E}_0/\sqrt{2}$, we find a
microwave trap depth for Cs of $\DE_{Cs} = -1.0$ mK, which is
easily sufficient to hold all atoms collected in a standard MOT.
Polarizabilities for all alkalis are similar. Atoms can be loaded
into the microwave trap by rapidly turning on the microwave field
after the MOT is loaded; since the volume of the microwave trap is
much larger than a typical MOT dimension, transfer should be very
efficient.

In this context, it is interesting to consider the elastic
cross-section between an alkali atom such as Cs, and a molecule in
the trap. This cross-section receives a contribution from the fact
that the alkali atoms are also slightly polarized in the microwave
field, with an rms expectation value of the atomic dipole moment
of $\left\langle{\mu}\right\rangle_{atom} = \alpha_0
\mathcal{E}_0/\sqrt{2}$. Taking into account only the
dipole-dipole interaction between atom and molecule, it is
possible to calculate the cross-section by a simple modification
of Eq.(\ref{crosssection}) to take into account scattering by
unequal dipoles.  This should represent at least a lower bound on
the atom-molecule elastic cross-section, which could in principle
be significantly enhanced by shorter-range effects (such as the
direct interaction of the dipolar field of the molecule with the
polarizability of the atom) that are not considered here.  For Cs
under our nominal conditions, we find
$\left\langle{\mu}\right\rangle_{Cs} = \alpha_0
\mathcal{E}_0/\sqrt{2} = 3 \times 10^{-3}$ D and a Cs-SrO cross
section, due to dipole-dipole interactions, of $\sigma_{Cs-SrO}
\approx 10^{-12}$ cm$^2$ at $T \sim 1$ K.  This is large enough to
enable rapid sympathetic cooling with typical atomic densities
achieved in a MOT (see e.g. Ref. \cite{ourRbCspapers}).

\section{Conclusions}
\label{sec5} We have argued that microwave-based traps for polar
molecules have many attractive properties.  In particular, using
this technology it appears possible to create large, deep traps
for absolute ground state molecules.  Collisions between molecules
in these traps should have very large elastic cross-sections and
no two-body inelastic losses, opening the realistic possibility
for evaporative cooling of molecules.  Our discussion has stressed
the design of such a trap for diatomic molecules in rigid rotor
states, using an open trap geometry.  With minor modifications
from our nominal design parameters, this trap would be well-suited
to a wide variety of species; in particular, it works well for
molecules with rotational constants $B_e$ in the range of roughly
0.05 -0.5 cm$^{-1}$ and intrinsic dipole moments $\mu$ of at least
few Debye.  This covers many diatomics with the lightest
constituent atom in the first two complete rows of the periodic
table (Li - Cl).  Additional design considerations would be
required to optimize the microwave trap for hydrides, or for
molecules with inversion or lambda-doublet structures; however, we
think it is likely that a similar type of technology could prove
useful in these cases as well.

There also remain a number of interesting questions concerning the
operation of such traps.  In particular, possible inelastic losses
due to three-body processes, particularly those resulting in
chemical reactions, will be interesting to study in this
strongly-interacting system.  In addition, the dynamics of
evaporative cooling in such traps may be very different from the
well-studied cases in atomic physics, as the elastic
cross-sections become so large that the dipolar gas has a
non-negligible viscosity.  We look forward to future studies of
such issues.  As mentioned above, we have experiments underway to
load the strongly polar species SrO into such a trap, where we
anticipate many interesting developments.

We gratefully acknowledge support from NSF grant DMR-0325580, the
David and Lucile Packard Foundation, and the W.M. Keck Foundation.

%
%
%
%
%

\end{document}